\documentclass[seceq]{ptptex}

\usepackage{graphicx}
\usepackage{wrapft}

\title{
Quartetting in Nuclear Matter%
}

\author{
Peter \textsc{Schuck}$^{1,2}$,
Takaaki \textsc{Sogo}$^{1}$,
and
Gerd \textsc{R\"opke}$^{3}$
}

\inst{
$^{1}$
Institut de Physique Nucl\'eaire, CNRS-IN2P3 
and Universit\'e Paris-Sud, 91406 Orsay Cedex, France\\
$^{2}$
Laboratoire de Physique et Mod\'elisation 
des Milieux Condens\'es, CNRS and Universit\'e
Joseph Fourier, 25 Avenue des Martyrs, 
Bo\^ite Postale 166, F-38042 Grenoble Cedex 9, France\\
$^{3}$
Universit\"at Rostock, FB Physik, Universit\"atplatz,
D-18051 Rostock, Germany}

\abst{
A general theory for the condensation of strongly bound quartets in infinite 
nuclear matter is presented. Critical temperatures for symmetric and 
asymmetric nuclear matter are evaluated. A fully nonlinear theory for the 
quartet order parameter, based on an analogy of the Gorkov approach to pairing, 
is presented and solved. The strong qualitative difference with pairing is 
pointed out.
}

\begin{document}

\maketitle

\section{Introduction}

Since about one decade, there has been a renewed and strong interest in gas 
like $\alpha$ particle states in nuclei, as, e.g. the famous Hoyle state 
in $^{12}$C. 
This stems essentially from the fact that in 2001 the prediction was 
made that besides the Hoyle state with three $\alpha$'s, other similar states 
with more $\alpha$'s can exist and that those states can be considered as 
being of $\alpha$ particle condensate type \cite{Toh01}. Of course, in 
finite systems, 
like nuclei, neither pairing nor quartetting can show phase locking as in 
macroscopic systems. They can be considered as precursors to condensation 
phenomena in macroscopic systems, like neutron or nuclear matter. The proposed 
THSR wave function \cite{Toh01} for the description of the $\alpha$ gas 
states is very 
successful for nuclei. It is a number conserving ansatz, in spirit similar 
to a number projected BCS wave function, which cannot be 
extended to the infinite matter case. We, therefore, will discuss in this 
article how $\alpha$ particle condensation can be described for the case of 
infinite nuclear matter. This may be relevant for compact stellar objects. 
However, the general theory may also serve for other systems where quartet 
correlations are important as, e.g., in a gas of excitons where bi-excitons 
may be formed. Also in the physics of cold atoms the possibility of trapping 
four different fermions and, thus, of quartet condensation, may exist 
in the future. 
The paper is organised as follows. In \S\ref{sec-2}, we will describe 
how to determine the critical temperature in symmetric and asymmetric matter. 
In \S\ref{sec-3}, a general formula for the quartet order parameter will be 
presented and applied to symmetric nuclear matter using a Gorkov type of 
formalism. Finally in \S\ref{sec-4} we give some perspectives and conclude.

\section{\label{sec-2}
Critical temperature of quartetting and $\alpha$ particle condensation}

In this section, we will discuss the critical temperature of quartet 
condensation in symmetric and asymmetric nuclear matter. A comparison with 
the pairing case will also be given. For this, we start out to 
establish an in medium four body equation.

In medium four particle correlation for example appears if one adds
an $\alpha$-particle on top of a double magic nucleus such as $^{208}$Pb,
or in semiconductors where in the gas of excitons, i.e. $p-h$ bound
states, may appear bi-excitons, i.e. bound state of two $p-h$ pairs.
The effective wave equation contains, in mean field approximation, the 
Hartree-Fock self-energy shift of the single-particle energies as well as 
the Pauli blocking of the interaction. 
We consider for the derivation the Equation of Motion Method (EMM), well 
known from the Random Phase Approximation (RPA) \cite{Rin80}.
\begin{equation}
\label{com4}
\langle[\delta Q,[H,\delta Q^{\dagger}_{\alpha}]]\rangle
=E_{\rho}\langle[\delta Q,Q^{\dagger}_{\alpha}]\rangle,
\end{equation}
where
\begin{equation}
Q^{\dagger}_{\alpha}=\sqrt{\frac{1}{4!}}\sum_{1234}\psi_{4,\alpha}(1234)
a^{\dagger}_1a^{\dagger}_2a^{\dagger}_3a^{\dagger}_4,
\end{equation}
is the quartet creator. Of course, we also could consider analogously
a quartet destructor.
Considering a fermion Hamiltonian with two body interaction, we evaluate
the double commutator in (\ref{com4}) and perform the expectation value
with the Hartree-Fock ground state, which is, as we know, the standard
procedure to obtain RPA equations in linearised form. The result is
\cite{Rop98,Bey00}
\begin{eqnarray}
\label{EWE}
&&
(E_{\rho}-\epsilon_1-\epsilon_2-\epsilon_3-\epsilon_4)\psi_{4,\rho}(1234)
\nonumber \\
&=&
[1-f_1-f_2]\bar{V}_{121'2'}\psi_{4,\rho}(1'2'34)+{\rm permutations},
\end{eqnarray}
where $\epsilon_1$  is the single particle energy, i.e. the kinetic energy plus 
HF shift.
A similar equation can be obtained for three particles, i.e. $A=3$. 
This equation has exactly the same structure as a free four body
Schr\"odinger equation with the only difference that the two body matrix
elements $\bar{V}_{iji'j'}$ are premultiplied with the phase space
factor $\bar{f}_i\bar{f}_j-f_if_j=1-f_i-f_j$ where the $f_i$ are
the Fermi-Dirac functions, and $\bar{f}_i = 1 - f_i$. The Fermi function is 
equal to the step function at $T=0$,
i.e. $f_i=\Theta(\mu-\epsilon_i)$, but immediately generalisable to finite
temperature with $f_i=[1+e^{(\epsilon_i-\mu)/T}]^{-1}$, a result which could have
been derived employing Matsubara Green functions.
In (\ref{EWE}) the $\epsilon_i$ are as usual the HF single particle energies
and $\bar{V}_{iji'j'}=\langle ij|V|i'j'\rangle-\langle ij|V|j'i'\rangle$ is the
antisymmetrized matrix element of the two body interaction.
We realise that the phase space factor in (\ref{EWE}) is exactly the same
as in the two particle RPA.\cite{Rin80} It turns out to be a general rule
in deriving higher $pp$ or $ph$ RPA's that the $pp$ and $ph$ matrix
elements are always premultiplied with the same phase space factors
as in the corresponding standard $pp$ or $ph$ RPA's.

We are interested in an example of nuclear physics where the $\alpha$-particle
constitutes a particularly strongly bound cluster of four nucleons.
One can ask the question how, for a fixed temperature, the binding energy
of the $\alpha$-particle varies with increasing temperature.

\begin{wrapfigure}{r}{6.6cm}
\includegraphics[width=60mm]{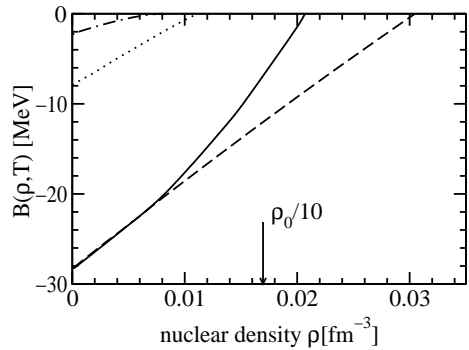}
\caption{
Shift of binding energy of the light clusters 
($d$ - dash dotted, $t/h$ - dotted, and $\alpha$ - dashed: perturbation theory,
full line:~non-perturbative Faddeev-Yakubovski equation) in symmetric nuclear 
matter as a function of density for given temperature 
$T = 10$ MeV.\cite{Bey00}}
\label{shifts}
\end{wrapfigure}

The effective wave equation has been solved using separable potentials
for $A=2$ by integration. For $A=3,4$ we can use a 
{\it Faddeev approach} \cite{Sog09}.
The shifts of binding energy can 
also be calculated approximately via perturbation theory.  
In Fig.~\ref{shifts} we show the shift of the binding energy of the light 
clusters (deuteron ($d$), triton ($t$), helion ($h$) and $\alpha$) in symmetric 
nuclear matter as a function 
of density for temperature $T$ = 10 MeV~\cite{Bey00}.

It is found that the cluster binding energy decreases with increasing density.
Finally, at the {\it Mott density} $\rho_{A,n,P}^{\rm Mott}(T)$ the bound 
state is dissolved. The clusters are not present at higher densities,  
merging into the nucleonic medium.  
It is found that the $\alpha$ particle already dissolves at a density  
$\rho_\alpha^{\rm Mott} \approx \rho_0/10$,  
see Fig.~\ref{shifts}.   For a given cluster type characterized 
by $A,n$, we can also introduce the Mott momentum 
$P^{\rm Mott}_{A,n}(\rho,T)$ in terms of the ambient temperature $T$ and 
nucleon density $\rho$, such that the bound states exist only for 
$P \ge P^{\rm Mott}_{A,n}(\rho,T)$.  We do not present an example here, 
but it is intuitively clear that a cluster with high c.o.m. momentum with 
respect to the medium is less
affected by the Pauli principle than a cluster at rest.

Since Bose condensation only is of relevance for deuterons ($d$) 
and $\alpha$'s, and the 
fraction of $d$, tritons ($t$) and helions ($h$) becomes low compared with 
that of $\alpha$'s with 
increasing density, we can neglect the contribution of the latters 
to the equation 
of state. Consequently, if we further neglect the contribution of the 
four-particle scattering phase shifts in the different channels, we can now 
construct an equation of state 
$\rho(T, \mu) =\rho^{\rm free}(T, \mu) + \rho^{{\rm bound}, d}
(T, \mu) +\rho^{{\rm bound}, \alpha}(T, \mu)$ such that 
$\alpha$-particles determine the behavior of symmetric nuclear matter at 
densities below $\rho_\alpha^{\rm Mott}$ and temperatures below the 
binding energy per nucleon of the $\alpha$-particle. The formation of 
deuteron clusters alone gives an incorrect description because the 
deuteron binding energy is small, and, thus, the abundance of $d$-clusters 
is small compared with that of $\alpha$-clusters. In the low density region 
of the phase diagram, $\alpha$-matter emerges as an adequate model for 
describing the nuclear-matter equation of state.

With increasing density, the medium modifications -- especially Pauli 
blocking -- will lead to a deviation of the critical temperature 
$T_c(\rho)$ from that of an ideal Bose gas of $\alpha$-particles 
(the analogous situation holds for deuteron clusters, i.e., in the 
isospin-singlet channel).

Symmetric nuclear matter is characterized by the equality of the proton 
and neutron chemical potentials, i.e., $\mu_p=\mu_n=\mu$. Then an extended 
Thouless condition based on the relation for the four-body T-matrix 
(in principle equivalent to (\ref{EWE}) at eigenvalue 4$\mu$)
\begin{eqnarray}
{\rm T}_4(1234,1''2''3''4'', 4 \mu)
&=& \sum_{1'2'3'4'} \Biggl[
  \frac{{V_{12,1'2'}}[1-f_1-f_2] }{ 4
    \mu - \epsilon_1-\epsilon_2-\epsilon_3-\epsilon_4 }
\delta(3,3')\delta(4,4')
+ {\rm cycl.} \Biggr]
\nonumber \\
&&\times
{\rm T}_4(1'2'3'4',1''2''3''4'', 4 \mu)
\label{eq:4body_T_matrix}
\end{eqnarray}
serves to determine the onset of Bose condensation of $\alpha$-like clusters, 
noting that the existence of a solution of this relation signals a divergence 
of the four-particle correlation function. An approximate solution has been 
obtained by a variational approach, in which the wave function is taken as 
Gaussian incorporating the correct solution for the two-particle 
problem~\cite{Rop98}.

\begin{wrapfigure}{r}{6.6cm}
\includegraphics[width=60mm]{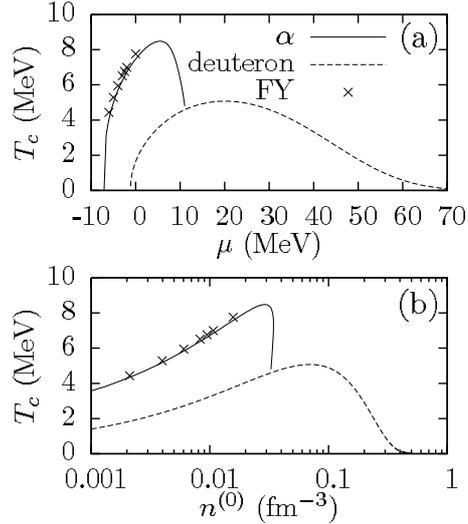}
\caption{Critical temperature of alpha and deuteron condensations 
as functions of (a)~chemical potential and (b)~density of free nucleon
{~\cite{Sog09}}. Crosses ($\times$) correspond to calculation 
of Eq.~(\ref{EWE}) with the Malfliet-Tjon interaction (MT I-III) using the 
Faddeev-Yakubovski method.}
\label{fig2}
\end{wrapfigure}

On the other hand, Eq.~(\ref{eq:4body_T_matrix}), respectively 
(\ref{EWE}) at eigenvalue $4\mu$, has also been solved numerically 
exactly by the Faddeev-Yakubovsky method employing the Malfliet-Tjon 
force.\cite{Mal69}  The results for the critical temperature 
of $\alpha$-condensation is presented in Fig.~\ref{fig2} as a function of the 
chemical potential {$\mu$}. The exact solution could only be 
obtained for negative {$\mu$}, i.e.~when there exists a bound 
cluster. It is, therefore, important to try yet another approximate solution 
of the in-medium four-body equation. Since the $\alpha$-particle is strongly 
bound, we make a momentum projected mean field ansatz for the quartet wave 
function \cite{Rin80,Kam05}
\begin{eqnarray}
\Psi_{1234}
&=&
(2\pi)^3 \delta^{(3)}({\bf k}_1 +{\bf k}_2 + {\bf k}_3 + 
{\bf k}_4) 
\nonumber \\
&&\times\prod_{i=1}^4\varphi({\bf k}_i)\chi^{ST},
\label{eq3}
\end{eqnarray}
where $\chi^{ST}$ is the spin-isospin function which we suppose to be the 
one of a scalar ($S=T=0$). We will not further mention it from now on. 
We work in momentum space and $\varphi({\bf k})$ is the as-yet unknown 
single particle $0S$ wave function. In position space, this leads to the 
usual formula \cite{Rin80} 
$\Psi_{1234} \rightarrow \int d^3R \prod_{i=1}^4 \tilde\varphi({\bf r}_i - 
{\bf R})$ where $\tilde\varphi({\bf r}_i)$ is the Fourier transform of 
$\varphi({\bf k}_i)$. If we take for $\varphi({\bf k}_i)$ a Gaussian shape, 
this gives: 
$\Psi_{1234} \rightarrow \exp[-c\sum_{1\leq i<k \leq 4} ({\bf r}_i - 
{\bf r}_k)^2]$ which is the translationally invariant ansatz often used 
to describe $\alpha$-clusters in nuclei. For instance, it is also employed 
in the $\alpha$-particle condensate wave function of Tohsaki, Horiuchi, Schuck,
R\"opke (THSR) in Ref.~\citen{Toh01}.

Inserting the ansatz (\ref{eq3}) into (\ref{EWE}) and integrating over 
superfluous variables, or minimizing the energy, we arrive at a Hartree-Fock 
type of equation for the single particle $0S$ wave function 
$\varphi(k)=\varphi(|{\bf k}|)$ which can be solved. However, for a general 
two body force 
${V_{{\bf k}_1 {\bf k}_2, {\bf k}'_1 {\bf k}'_2}}$, 
the equation to be solved is still rather complicated. We, therefore, 
proceed to the last simplification and replace the two body force by a 
unique separable one, that is
\begin{equation}
{V_{{\bf k}_1 {\bf k}_2, {\bf k}'_1 {\bf k}'_2}} = 
\lambda e^{-k^2/k_0^2}e^{-k'^2/k_0^2} (2\pi)^3\delta^{(3)}({\bf K}-{\bf K}'),
\label{eq05}
\end{equation}
where 
${\bf k}=({\bf k}_1-{\bf k}_2)/2$, ${\bf k}'=({\bf k}_1'-{\bf k}_2')/2$, 
${\bf K}={\bf k}_1+{\bf k}_2$, and ${\bf K}'={\bf k}_1'+{\bf k}_2'$. 
This means that we take a spin-isospin averaged two body interaction and 
disregard that in principle the force may be somewhat different in the 
$S,T = 0, 1$ or $1, 0$ channels. It is important to remark that for a mean 
field solution the interaction only can be an effective one, very different 
from a bare nucleon-nucleon force. This is contrary to the usual gap equation 
for pairs, to be considered below, where, at least in the nuclear context, 
a bare force can be used as a reasonable first approximation.

We are now ready to study the solution of (\ref{EWE})
and (\ref{eq:4body_T_matrix}) 
for the critical temperature $T_c^{\alpha}$, defined by the point where 
the eigenvalue equals $4\mu$. For later comparison, the deuteron (pair) 
wave function at the critical temperature is also deduced from 
the Thouless criterion for the onset of pairing with
(\ref{eq05}) to be
\begin{equation}
\phi(k)= -\frac{1-2f(\varepsilon(k))}{k^2/m-2\mu}\lambda e^{-k^2/k_0^2}
\int \frac{d^3k'}{(2\pi)^3} e^{-k^2/k_0^2} \phi(k'),
\label{eq08}
\end{equation}
where $\phi(k)$ is the relative wave function of two particles given by 
$\Psi_{12} \to {\phi(|\frac{{\bf k}_1-{\bf k}_2}{2}|)}$ 
${\delta^{(3)}({\bf k}_1+{\bf k}_2)}$, and $\varepsilon(k)=k^2/(2m)-\mu$. 
We also neglected the momentum dependence of the Hartree-Fock mean field 
shift in {Eq.~(\ref{eq08})}. It, therefore, can be 
incorporated into the chemical potential $\mu$. With Eq.~(\ref{eq08}), 
the critical temperature of pair condensation is obtained from the following 
equation:
\begin{equation}
1=-\lambda \int \frac{d^3k}{(2\pi)^3}
\frac{1-2f(\varepsilon(k))}{k^2/m-2\mu} e^{-2k^2/k_0^2}.
\label{eq010}
\end{equation}

In order to determine the critical temperature for $\alpha$-particle 
condensation, we have to adjust the temperature so that the eigenvalue 
of (\ref{EWE}) and (\ref{eq:4body_T_matrix}) equals $4\mu$. The result 
is shown in Fig. \ref{fig2}(a). In order to get an idea how this converts 
into a density dependence, we use for the moment the free gas relation 
between the density $n^{(0)}$ of uncorrelated nucleons and the chemical 
potential
\begin{equation}
n^{(0)}=4\int \frac{d^3k}{(2\pi)^3} f(\varepsilon).
\label{eq-density}
\end{equation}
We are well aware of the fact that this is a relatively gross simplification, 
for instance at the lowest densities, and we intend to generalize our theory 
in the future so that correlations are included into the density. This may be 
done along the work of Nozi\`eres and Schmitt-Rink \cite{Noz85}. 
The two open constants $\lambda$ and $k_0$ in 
Eq. (\ref{eq05}) are determined so that binding energy ($-28.3$ MeV) and 
radius ($1.71$ fm) of the free ($f_i=0$) $\alpha$-particle come out right. 
The adjusted parameter values are: $\lambda=-992$ MeV fm$^{3}$, and 
{$k_0=1.43$} fm$^{-1}$. The results of the calculation are 
shown in Fig.~\ref{fig2}.

In Fig.~\ref{fig2}, the maximum of critical temperature 
$T^{\alpha}_{c, {\rm max}}$ is at $\mu=5.5$ MeV, and the 
$\alpha$-condensation can exist up to  $\mu_{\rm max}=11$ MeV.  
It is very remarkable that the results obtained with (\ref{eq3}) 
for $T_c^{\alpha}$ very well agree with the exact solution of (\ref{EWE}) 
and (\ref{eq:4body_T_matrix}) using the Malfliet-Tjon interaction 
(MT I-III)~\cite{Mal69}
with the Faddeev-Yakubovski method also shown by 
crosses in Fig.~\ref{fig2} (the numerical solution only could be obtained 
for negative values of $\mu$). This indicates that $T_c^{\alpha}$ is 
essentially determined by the Pauli blocking factors.

In Fig.~\ref{fig2} we also show the critical temperature for deuteron 
condensation derived from Eq.~(\ref{eq010}). In this case, the bare force 
is adjusted with $\lambda= -1305$ MeV fm$^3$ and $k_0 = 1.46$ fm$^{-1}$ 
to get experimental energy ($-2.2$ MeV) and radius ($1.95$ fm) of the 
deuteron. It is seen that at higher densities deuteron condensation wins 
over the one of $\alpha$-particles. The latter breaks down rather abruptly 
at a critical positive value of the chemical potential. Roughly speaking, 
this corresponds to the point where the $\alpha$-particles start to 
overlap substantially. 
This behavior stems from the fact that Fermi-Dirac distributions in the 
four body case, see (\ref{eq:4body_T_matrix}), can never become 
step-like, as in the two body case, even not at zero temperature, since 
the pairs in an $\alpha$-particle are always in motion. Therefore, 
no threshold effect occurs as with pairing for Cooper pairs at rest. 
As a consequence, $\alpha$-condensation generally only exists as a BEC 
phase and the weak coupling regime is absent.

 An important consequence of this study is that at the lowest temperatures, 
Bose-Einstein condensation occurs for $\alpha$ particles rather than for 
deuterons.  As the density increases within the low-temperature regime, 
the chemical potential $\mu$ first reaches $-7$ MeV, where the $\alpha$'s 
Bose-condense.  By contrast, Bose condensation of deuterons would not occur 
until $\mu$ rises to $-1.1$ MeV.

The {\it ``quartetting''} transition temperature sharply drops as the rising 
density approaches the critical Mott value at which the four-body bound 
states disappear.  At that point, pair formation in the isospin-singlet 
deuteron-like channel comes into play, and a deuteron condensate will exist 
below the critical temperature for BCS pairing up to densities above the 
nuclear-matter saturation density $\rho_0$, as described in the previous 
Section. Of course, also the well known and studied isovector {\it n-n} and 
{\it p-p} pairing develops. 
Usually, $p$-$n$ pairing in the isoscalar channel is stronger.
However, the competition between the different pairing channels
is not well known.
The critical density at which 
the $\alpha$ condensate disappears is estimated to be $\rho_0/3$. 
Therefore, $\alpha$-particle condensation primarily only exists in the 
Bose-Einstein-Condensed (BEC) phase and there does not seem to exist a 
phase where the quartets acquire a large extension as Cooper pairs do in 
the weak coupling regime.  However, the variational approaches of 
Ref.~\citen{Rop98} and of Eq.~(\ref{eq3}) on which this conclusion is 
based represent only first attempts at the description of the transition 
from quartetting to pairing.  The detailed nature of this fascinating 
transition remains to be clarified. Many different questions arise in 
relation to the possible physical occurrence and experimental manifestations 
of quartetting: Can we observe the hypothetical ``$\alpha$ condensate'' 
in nature?  What about thermodynamic stability?  What happens with 
quartetting in asymmetric nuclear matter?  Are more complex quantum 
condensates possible?  What is their relevance for finite nuclei?  
As discussed, the special type of microscopic quantum correlations 
associated with quartetting may be important in nuclei, its role in these 
finite inhomogeneous systems being similar to that of pairing. 
\cite{Toh01}.

On the other hand, if at all, $\alpha$-condensation in compact star occurs 
at strongly asymmetric matter. It is, therefore, important to generalize the 
above study for symmetric nuclear matter to the asymmetric case. This can be 
done straightforwardly, again using our momentum projected mean field ansatz 
(\ref{eq3}) generalized to the asymmetric case. This implies to introduce 
two chemical potentials, one for neutrons and for protons. We also have to 
distinguish two single particle wave functions in our product ansatz which 
now reads \cite{Sog10}
\begin{eqnarray}
\psi_{1234}
\to
\varphi_p(k_1)\varphi_p(k_2)\varphi_n(k_3)\varphi_n(k_4)
\chi_0
(2\pi)^3\delta({\bf k}_1+{\bf k}_2+{\bf k}_3+{\bf k}_4)
\label{eq-phfwf}
\end{eqnarray}
\begin{wrapfigure}{r}{6.6cm} 
\includegraphics[width=60mm]{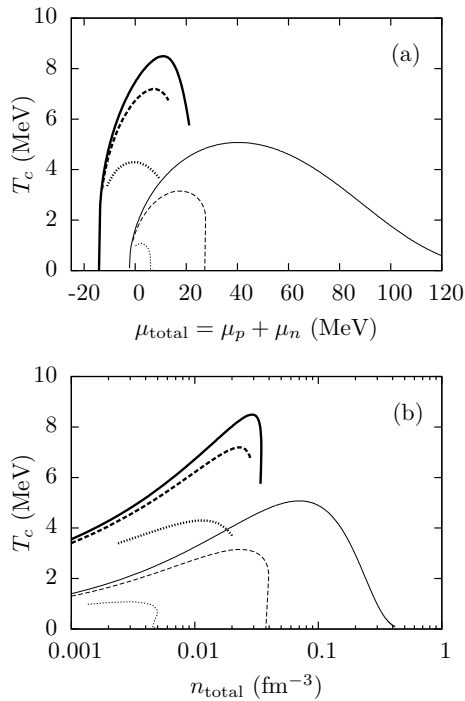}
\caption{Critical temperature as functions of the total chemical potential 
$\mu_{\rm total}=\mu_p+\mu_n$ (top) and the total free density 
$n_{\rm total}$ (bottom).\cite{Sog10} Thick (thin) 
lines are for $\alpha$-particle (deuteron). Solid, dashed, and dotted 
lines are respectively for $\delta=0.0$, $\delta=0.5$, and $\delta=0.9$, 
where the density ratio $\delta$ is in  Eq.~(\ref{eq-densityratio}).}
\label{fig-ntotalvstc}
\end{wrapfigure}
where $\varphi_{\tau}(k_i)=\varphi_{\tau}(|{\bf k}_i|)$ is the $s$-wave 
single particle wave functions for protons ($\tau=p$) and neutrons ($\tau=n$), 
respectively.  $\chi_0$ is the spin-isospin singlet wave function. 
This now leads to two coupled equations of the Hartree-Fock type for 
$\varphi_n$ and $\varphi_p$. For the force we use the same as in the symmetric 
case.

Fig.~\ref{fig-ntotalvstc}(a) shows the critical temperature of $\alpha$ 
condensation as a function of the total chemical potential 
$\mu_{\rm total}=\mu_p+\mu_n$. We see that $T_c$ decreases as the asymmetry, 
given by the parameter
\begin{eqnarray}
\delta=\frac{n_n-n_p}{n_n+n_p},
\label{eq-densityratio}
\end{eqnarray}
increases. This is in analogy with the deuteron case (also shown) which 
already had been treated in Refs.~\citen{Alm93,Lom01}. 
On the other hand, in Fig.~\ref{fig-ntotalvstc}(b), it is also interesting 
to show $T_c$ as a function of the free density which is
\begin{eqnarray}
n^{(0)}_{\rm total}&=&n^{(0)}_p+n^{(0)}_n
\label{eq-totaldensity}\\
n^{(0)}_{p,n}&=&2\int \frac{d^3k}{(2\pi)^3}f_{p,n}(k) 
\end{eqnarray}
where the factor two in front of the integral comes from the spin degeneracy, 
and $f_{p,n}(k)=[1+\exp(k^2/2m-\mu_{p,n})]^{-1}$. 
It should be emphasized, however, that in the above relation between 
density and chemical potential, the free gas relation is used and 
correlations in the density have been neglected. In this sense the 
dependence of $T_c$ on density only is indicative, more valid at the 
higher density side. The very low density part where the correlations 
play a more important role, will be treated in the future. For instance, it 
will be important to recover the the critical temperature corresponding to an 
ideal Bose gas in that limit. Techniques similar to the one of Nozi\`eres and 
Schmitt-Rink may be employed \cite{Noz85}.
It should, however, be stressed that the dependence of $T_c$ on the chemical 
potential as in Fig.~\ref{fig-ntotalvstc}(a), stays unaltered. It is only the 
relation between the chemical potential and the (correlated) density which 
changes.

The fact that for more asymmetric matter the transition temperature decreases, 
is natural, since as the Fermi levels become more and more unequal, 
the proton-neutron correlations will be suppressed. For small $\delta$'s, 
i.e., close to the symmetric case, $\alpha$ condensation (quartetting) breaks 
down at smaller density (smaller chemical potential) than deuteron 
condensation (pairing). This effect has already been discussed above 
for symmetric nuclear matter {~\cite{Sog09}}. 
For $\delta$'s close to one, i.e. strong asymmetries,  the behavior is 
opposite, 
i.e., deuteron condensation breaks down at smaller densities than $\alpha$ 
condensation, because the small binding energy of the deuteron can not 
compensate the difference of the chemical potentials.

More precisely, for small $\delta$'s, the deuteron with zero center of 
mass momentum is only weakly influenced by the density or the total 
chemical potential as can be seen in Fig.~\ref{fig-ntotalvstc}. However, 
as $\delta$ increases, the different chemical potentials for protons and 
neutrons very much hinders the formation of proton-neutron Cooper pairs 
in the isoscalar channel for rather obvious reasons. The point to make here 
is that because of the much stronger binding per particle of the 
$\alpha$-particle, the latter is much less influenced by the increasing 
difference of the chemical potentials. For the strong asymmetry $\delta=0.9$ 
in Fig.~\ref{fig-ntotalvstc} then finally $\alpha$-particle condensation 
can exist up to $n_{\rm total}=0.02$~fm$^{-3}$ ($\mu_{\rm total}=9.3$~MeV), 
while the deuteron condensation exists only up to 
$n_{\rm total}=0.005$~fm$^{-3}$ ($\mu_{\rm total}=6.0$~MeV).

Overall, the behavior of $T_c$ is more or less as expected. We should, 
however, remark that the critical temperature for $\alpha$-particle 
condensation stays quite high, even for the strongest asymmetry considered 
here, namely $\delta$ = 0.9. This may be of importance for the possibility 
of $\alpha$-particle condensation in neutron stars and supernovae 
explosions.

In conclusion the $\alpha$-particle (quartet) condensation was investigated 
in homogeneous symmetric nuclear matter as well as in asymmetric nuclear 
matter.  We found that the critical density at which the $\alpha$-particle 
condensate appears is estimated to be around ${\rho_0}/3$ in the symmetric 
nuclear matter, and the $\alpha$-particle condensation can occur only at low 
density. This result is consistent with the fact that the Hoyle state ($0^+_2$)
of $^{12}$C also has a very low density $\rho \sim \rho_0/3$. 
On the other hand, in asymmetric nuclear matter, the critical temperature 
$T_c$ for the $\alpha$-particle condensation was found to decrease with 
increasing asymmetry. However, $T_c$ stays relatively high for very strong 
asymmetries, a fact of importance in the astrophysical context. The asymmetry 
affects deuteron pairing more strongly than $\alpha$-particle condensation. 
Therefore, at high asymmetries, if at all, $\alpha$-particle condensate seems 
to dominate over pairing at all possible densities.

\section{\label{sec-3}`Gap' equation for the quartet order parameter}

For macroscopic $\alpha$ condensation it is, of course, not conceivable to 
work with a number projected $\alpha$ particle condensate wave function as 
we did when in finite nuclei only a couple of $\alpha$ particles 
were present.\cite{Toh01} 
We rather have to develop an analogous procedure to BCS theory but 
{generalized} for quartets. In principle a number non-conserving 
wave function of the type 
{$|\alpha\rangle = \exp[\sum_{1234}z_{1234}c_1^+c_2^+c_3^+c_4^+]|
{\rm vac}\rangle$} would be the ideal {generalization} of the BCS wave function
for the case of quartets. However, unfortunately, it is unknown so far 
(see, however, Ref.~\citen{Jem10}) 
how to treat such a complicated many 
body wave function mathematically in a reasonable way. So, we rather attack 
the problem from the other end, that is with a Gorkov type of approach, 
well known from pairing but here extended to the quartet case. 
Since, naturally, the formalism is complicated, we only will outline the 
main ideas and refer for details to the literature.

Actually one part of the problem is written down easily. Let us guide from a 
particular form of the gap equation in the case of pairing. We have at zero 
temperature 
\begin{eqnarray}
(\epsilon_1+\epsilon_2)\kappa_{12}
+ (1- \rho_1 - \rho_2)
\frac{1}{2}\sum_{1'2'} {\bar{V}_{121'2'}} 
\kappa_{1'2'} 
= 
2\mu\kappa_{12},
\label{eq:gap_eq_two_particles}
\end{eqnarray}
where $\kappa_{12} = \langle c_1c_2\rangle$ is the pairing tensor, 
$\rho_i = \langle c_i^+c_i \rangle
=\frac{1}{2}\left[1-\frac{\epsilon_i-\mu}{E_i}\right]$ 
with $E_i=\sqrt{(\epsilon_i-\mu)^2+\Delta_i^2}$ the quasi-particle energy
are the BCS occupation numbers, and 
{$\bar{V}_{121'2'}$ denotes the antisymmetrized matrix element of the 
two-body interaction.} The $\varepsilon_i$ are the usual mean field energies. 
Equation {(\ref{eq:gap_eq_two_particles})} is equivalent to the usual gap 
equation in the case of zero total momentum and opposite spin, i.e. in short 
hand: $2=\bar 1$ where the bar stands for 'time reversed conjugate'. 
The extension of {(\ref{eq:gap_eq_two_particles})} to the quartet case is 
formally written down without problem and can be derived with EMM, slightly 
extending (\ref{EWE}) to correlated occupation numbers at $T$=0.
\begin{eqnarray}
(\epsilon_{1234} - 4\mu)\kappa_{1234} 
&=& (1-\rho_1-\rho_2)
\frac{1}{2}\sum_{1'2'} {\bar{V}_{121'2'}} 
\kappa_{1'2'34} 
\nonumber \\
&+& (1-\rho_1-\rho_3)\frac{1}{2}\sum_{1'3'} {\bar{V}_{131'3'}} 
\kappa_{1'23'4} + {\rm all~permutations}.
\end{eqnarray}
with $\kappa_{1234} = \langle c_1c_2c_3c_4 \rangle $ the quartet order 
parameter.
This is formally the same equation as in Eq.~(\ref{EWE}) with, 
however, 
the Fermi-Dirac occupation numbers replaced by the zero temperature quartet 
correlated single particle occupation numbers, similar to the BCS case. 
For the quartet case, the crux lies in the determination of those occupation 
numbers. Let us again be guided by BCS theory or rather by the equivalent 
Gorkov approach.\cite{Fet71} In the latter, 
there are two coupled equations, one for the normal single particle 
Green's function (GF) and the other for the anomalous GF. Eliminating the 
one for the anomalous GF in inserting it into the first equation leads to 
a Dyson equation with a single particle mass operator \cite{Rin80},
\begin{wrapfigure}{r}{6.6cm} 
\includegraphics[width=60mm]{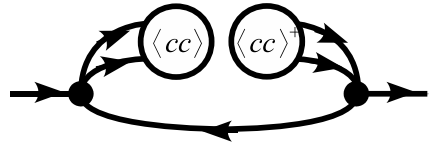}
\caption{Graphic representation of the BCS mass operator 
in Eq.~(\ref{eq-bcsmass})}
\label{fig-bcsmass}
\end{wrapfigure}
\begin{eqnarray}
&&
M^{\rm BCS}_{1;1'}(\omega) = \sum_{2}\frac{\Delta_{12}\Delta_{1'2}^*}
{\omega+\varepsilon_2}
\label{eq-bcsmass}
\end{eqnarray}
with
\begin{eqnarray}
\Delta_{12}=-\frac{1}{2}
\sum_{34} {\bar{V}_{12,34}} 
\kappa_{43}
\end{eqnarray}
This can be graphically represented as in Fig.~\ref{fig-bcsmass} where 
$\langle cc \rangle$ stands for the order parameter $\kappa_{12}$ and the 
dot for the two body interaction.

\begin{wrapfigure}{r}{6.6cm} 
\includegraphics[width=60mm]{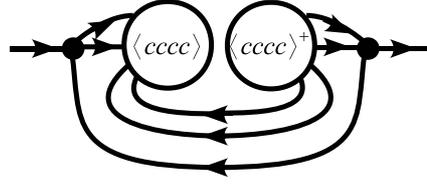}
\caption{Graphical representation of the approximate $\alpha$-BEC mass operator 
$M^{\rm quartet}$ of Eq.~(\ref{eq-Qmassoperator}). }
\label{fig-alphamassapp}
\end{wrapfigure}

The generalization to the quartet case is considerably more complicated 
but schematically the corresponding mass operator in the single particle 
Dyson equation can be represented graphically as in 
Fig.~\ref{fig-alphamassapp} with the quartet order parameter 
$\langle cccc\rangle$. 
This mass operator can again be derived employing 
the EMM. However, the details are rather lengthy and involved and, thus, we 
refrain from presenting this here, see \cite{Sog10}. Put aside this 
difficulty 
to derive a manageable 
expression for this 'quartet' s.p. mass operator, what immediately strikes 
is that instead of only one 'backward going line' with $(-{\bf p},-\sigma)$ 
as in the pairing case, we now have three backwards going lines. 
As a consequence, the three momenta ${\bf k}_1$, ${\bf k}_2$, ${\bf k}_3$ 
in these lines are only constrained so that their sum be equal to 
${\bf k}_1 + {\bf k}_2 + {\bf k}_3 = - {\bf p}$ in order that the total 
momentum of the order parameter be zero and, thus, the remaining 
freedom has to be summed over. This is in strong contrast to the pairing 
case where the single backward going line is constrained by momentum 
conservation to have momentum $-{\bf p}$. So, no internal summation 
occurs in the mass 
operator belonging to pairing. The consequence of this additional momentum 
summation in the mass operator for quartetting leads with respect to pairing 
to a completely different analytic structure of the mass operator in case of 
quartetting. This is best studied with the so-called three hole level density 
$g_{3h}(\omega)$ which is related to the imaginary part of the three hole 
Green's function 
${G^{3h}(k_1, k_2,k_3; \omega)} = 
({\bar f_1} {\bar f_2} {\bar f_3} + f_1f_2f_3)/(\omega + 
\varepsilon_{123})$ with $\varepsilon_{123}=\varepsilon_{1}+\varepsilon_{2}+
\varepsilon_{3}$ figuring in the mass operator, 
see Fig.~\ref{fig-alphamassapp} (we remind: ${\bar f}= 1-f$)
\begin{eqnarray}
g_{3h}(\omega) &=&-\int \frac{d^3k_1}{(2\pi)^3}\frac{d^3k_2}{(2\pi)^3}
\frac{d^3k_3}{(2\pi)^3} {\rm Im} G^{(3h)}(k_1,k_2,k_3;\omega+i\eta)
\nonumber \\
&=&\int \frac{d^3k_1}{(2\pi)^3}\frac{d^3k_2}{(2\pi)^3}\frac{d^3k_3}{(2\pi)^3}
(\bar f_1 \bar f_2 \bar f_3+ f_1 f_2 f_3)
\pi\delta(\omega+\varepsilon_1
                +\varepsilon_2
                +\varepsilon_3).
\label{eq-ld}
\end{eqnarray}

\begin{wrapfigure}{r}{6.6cm} 
\includegraphics[width=60mm]{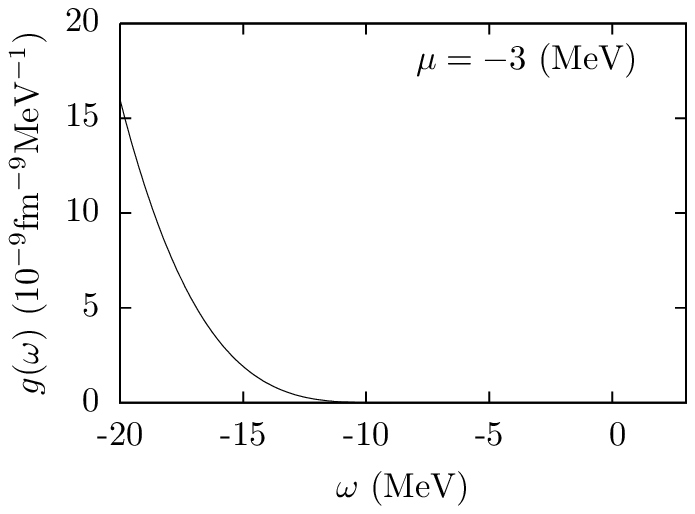}\\
\includegraphics[width=60mm]{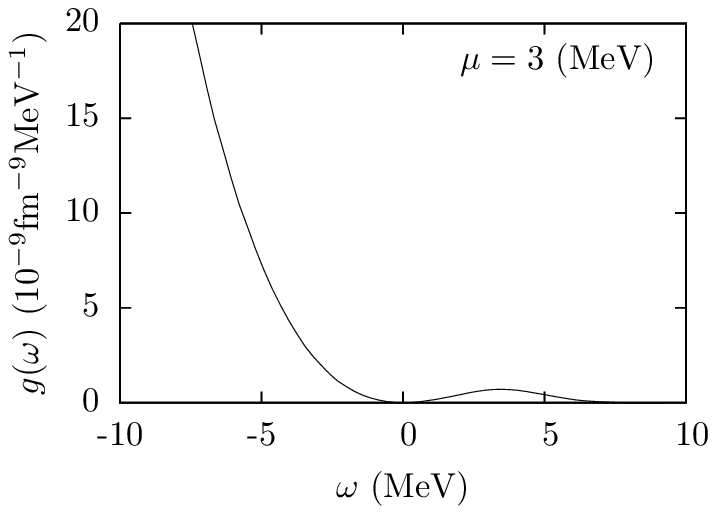}\\
\includegraphics[width=60mm]{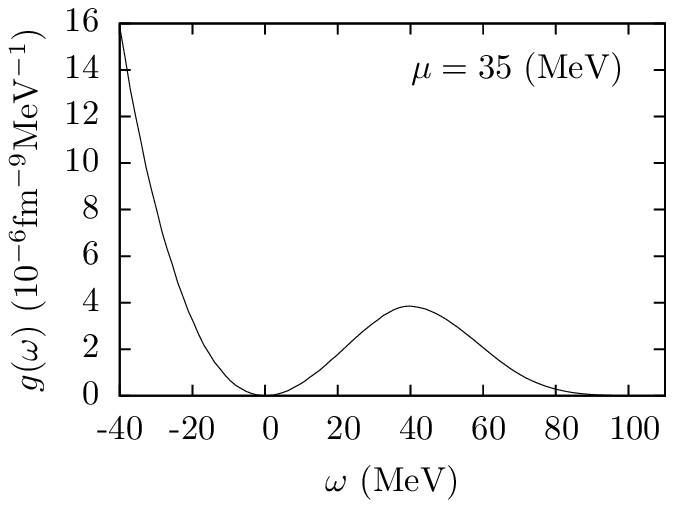}
\caption{
{$3h$} level densities defined in Eq.~(\ref{eq-ld}) for various values of 
the chemical potential $\mu$ at a zero temperature.\cite{Sog10} }
\label{fig-ld}
\end{wrapfigure}

In Fig.~\ref{fig-ld} we show the level density at zero temperature 
($f(\omega)=\theta(-\omega)$), where it is calculated with the proton mass 
$m=938.27$~MeV.\cite{Sog10} 
Two cases have to be considered, chemical potential 
$\mu$ positive or negative. In the latter case we have binding of the quartet. 
Let us first discuss the case $\mu>0$. We remark that in this case, the 
{$3h$} level density goes through zero at $\omega=0$, i.e., since we are 
measuring energies with respect to the chemical potential $\mu$, just in the 
region where the quartet correlations should appear. This is at strong 
variance with the pairing case where the {$1h$} level density, 
$g_{1h}(\omega)=\int \frac{d^3k}{(2\pi)^3} (\bar f_k + f_k)
\delta(\omega+\varepsilon(k)) = \int \frac{d^3k}{(2\pi)^3}
\delta(\omega+\varepsilon(k))$, does not feel any influence from the medium 
and, therefore, the corresponding level density varies (neglecting the mean 
field for the sake of the argument) like in free space with the square root  
of energy. In particular, this means that the level density is {\it finite} 
at the Fermi level. This is a dramatic difference with the quartet case and 
explains why Cooper pairs can strongly overlap whereas for quartets this is 
impossible as we will see below. We also would like to point out that the 
{$3h$} level density is just the mirror to the {$3p$} level density which has 
been discussed in Ref.~\citen{Bli86}.

For the case where $\mu<0$ where anyway the $f_i$'s are zero at $T$=0,  
there is nothing very special, besides the fact 
that the three hole level density only is non-vanishing for negative values 
of $\omega$ and that the 
upper boundary is given by $\omega = 3\mu$. Therefore, the level density of 
Eq.~(\ref{eq-ld}) is zero for $\omega>3\mu$. Therefore, in the BEC 
regime ($\mu < 0$), there is no marked difference between the pairing and 
quartetting cases.
 
With these preliminary but crucial considerations we now pass to the 
evaluation of the s.p. mass operator with quartet condensation. 
Its expression corresponding to Fig.~\ref{fig-ld} can be shown to be of the 
following form
\begin{eqnarray}
&&
M^{\rm quartet}_{1;1}(\omega)
\nonumber \\
&=&\sum_{234}
\frac{\tilde \Delta_{1234}(\bar f_2 \bar f_3 \bar f_4+ f_2 f_3 f_4)
\tilde \Delta_{1234}^*}{\omega+\varepsilon_{234}}
\nonumber \\
\label{eq-Qmassoperator}
\end{eqnarray}
with
\begin{eqnarray}
\tilde \Delta_{1234}
=
\sum_{1'2'3'4'}\frac{1}{2} {\bar{V}_{12,1'2'}} \delta_{33'}\delta_{44'}
\kappa_{1'2'3'4'}
\nonumber \\
\label{eq-tildedelta}
\end{eqnarray}
Again, comparing the quartet s.p. mass operator {(\ref{eq-Qmassoperator})} 
with the pairing one {(\ref{eq-bcsmass})}, we notice the presence of the 
phase space factors in the former case while in {Eq.~(\ref{eq-bcsmass})} 
they are absent. As already indicated above, this fact implies in the 
quartet case that only the Bose-Einstein condensation phase is born out 
whereas a 'BCS phase' (long coherence length) is absent, since the three hole 
level density is zero at 3$\mu$ due to the non vanishing phase space factors.

The complexity of the calculation in {Eq.~(\ref{eq-Qmassoperator})} 
is  much reduced using for the order parameter {${\langle cccc \rangle}$} 
our mean field ansatz projected on zero total momentum, as it was already 
very successfully employed with Eq.~(\ref{eq3}),
\begin{eqnarray}
&&
\langle c_1c_2c_3c_4 \rangle
\rightarrow
\phi_{{\bf k}_1 {\bf k}_2,{\bf k}_3 {\bf k}_4}\chi_0, 
\nonumber \\
&&
\phi_{{\bf k}_1 {\bf k}_2,{\bf k}_3 {\bf k}_4}
=
\varphi(|{\bf k}_1|)\varphi(|{\bf k}_2|)\varphi(|{\bf k}_3|)\varphi(|{\bf k}_4|)
(2\pi)^3\delta({\bf k}_1+{\bf k}_2+{\bf k}_3+{\bf k}_4),
\label{eq-PHF4bwf}
\end{eqnarray}
where $\chi_0$ is the spin-isospin singlet wave function.
It should be pointed out that this product ansatz with four identical $0S$ 
single particle wave functions is typical for a ground state configuration of 
the $\alpha$ particle. Excited configurations with wave functions of 
higher nodal structures may eventually be envisaged for other physical 
situations. We also would like to mention that the momentum conserving 
$\delta$ function induces strong correlations among the four particles 
and (\ref{eq-PHF4bwf}) is, therefore, a rather non trivial variational 
wave function.
 
For the two-body interaction of 
${{V}_{{\bf k}_1{\bf k}_2,{\bf k}_3 {\bf k}_4}}$ in Eq.~(\ref{eq-tildedelta}), 
we employ the same separable form (\ref{eq05}) as done already for the quartet 
critical temperature.
 
\begin{figure}
\begin{center}
\includegraphics[width=60mm]{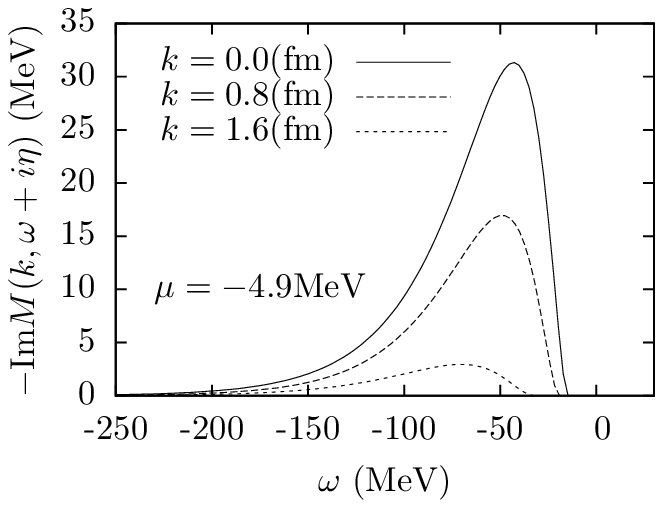}
\includegraphics[width=60mm]{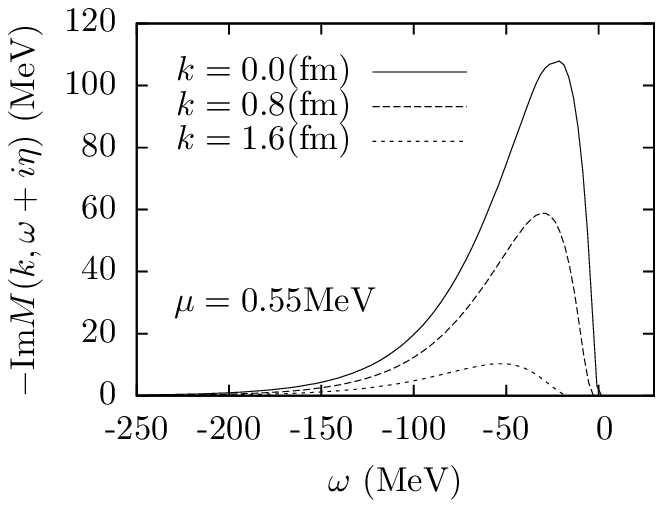}
\end{center}
\caption{\label{fig-immassoperator}
$-$Im$M^{\rm quartet}(k_1,\omega+i\eta)$ in Eq.~(\ref{eq-Qmassoperator}) 
as a function of $\omega$ for $\mu=-4.9$MeV (left) and for 
$\mu=0.55$MeV (right)  at  zero temperature.\cite{Sog10}}
\end{figure}
 
At first let us mention that in this pilot application of our selfconsistent 
quartet theory, we only will consider the zero temperature case. 
As a definite physical example, we will treat the case of nuclear physics 
with the particularly strongly bound quartet, the $\alpha$ particle. 
It should be pointed out, however, that if scaled appropriately all energies 
and lengths can be transformed to other physical systems. For the nuclear 
case it is convenient to measure energies in Fermi energies 
$\varepsilon_F = 35$~MeV and lengths in inverse Fermi momentum 
$k_F^{-1} = 1.35^{-1}$~fm.
 
\begin{figure}[t]
\begin{center}
\includegraphics[width=120mm]{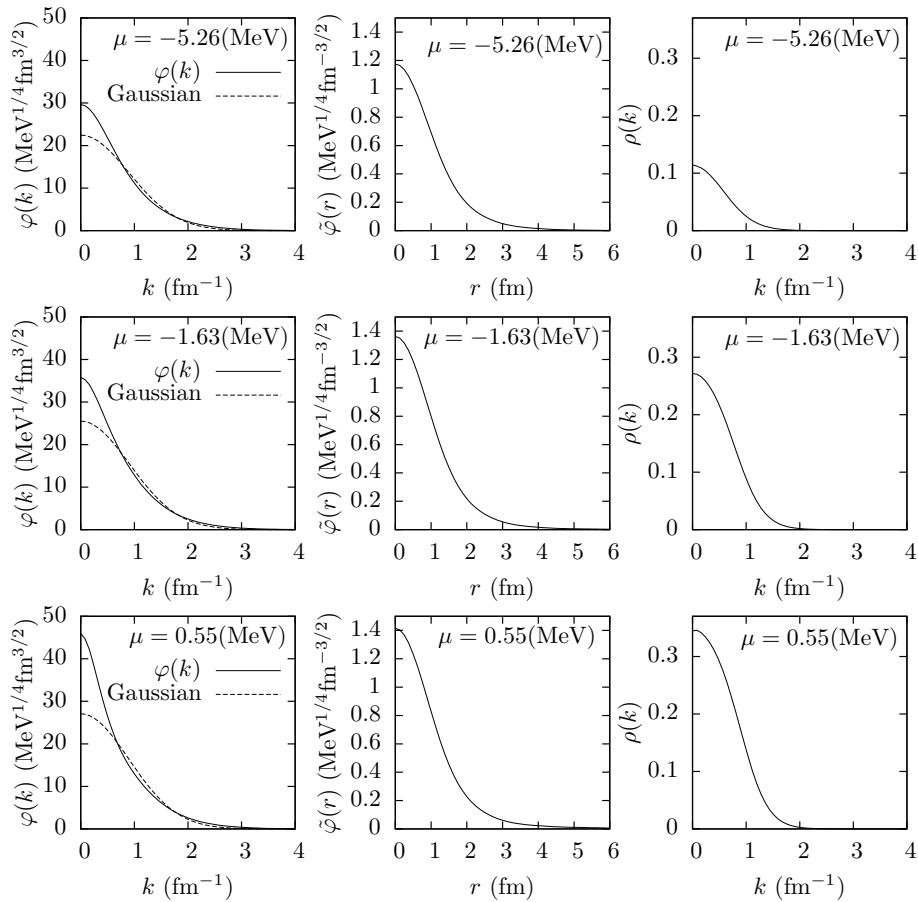}
\caption{\label{fig-spwf_1}
Single particle wave function $\varphi(k)$ in $k$-space (left), for 
$r$-space $\tilde \varphi(r)$ (middle), and occupation numbers (right) at 
$\mu=-5.26$ (top), $-1.63$ (middle) and $0.55$ (bottom). 
The $r$-space wave function $\tilde \varphi(r)$ is derived from the 
Fourier transform of $\varphi(k)$ by 
$\tilde \varphi(r)=\int d^3k e^{i{\bf k} \cdot {\bf r}}\varphi(k)/(2\pi)^3$. 
The dashed line in the left panels correspond to the Gaussian with same norm 
and {r.m.s.} momentum as $\varphi(k)$.\cite{Sog10}}
\end{center}
\end{figure}

We are now in a position to solve, as in the BCS case, the two coupled 
equations (\ref{eq:gap_eq_two_particles}) 
for the quartet order parameter and the single particle 
occupation numbers from the single particle Dyson equation with single 
particle self energy (\ref{eq-Qmassoperator}) with (\ref{eq-PHF4bwf}) 
selfconsistently.
The single particle wave functions and occupation numbers obtained from 
the above cycle are shown in Fig.~\ref{fig-spwf_1}. We also insert the 
Gaussian wave function with same r.m.s. momentum as the single particle 
wave function in the left figures in Fig.~\ref{fig-spwf_1}. 
As shown in Fig.~\ref{fig-spwf_1}, the single particle wave function is 
sharper than a Gaussian.

We could not obtain a convergent solution for $\mu>0.55$~MeV. 
This difficulty has precisely it's origin in the fact that the three hole 
level density goes through zero at 3$\mu>0$, just where the four body 
correlations should build up, as this was discussed above. 
In the r.h.s. panels of Fig.~\ref{fig-spwf_1} we also show the corresponding 
occupation numbers. We see that they are very small. 
However, they increase 
for increasing values of the chemical potential. For $\mu = 0.55$~MeV 
the maximum of the occupation still only attains 0.35 what is far away from 
the saturation value of one. What really happens for larger values of the 
chemical potential, remains unclear. 
 
\begin{figure}[t]
\begin{center}
\includegraphics[width=60mm]{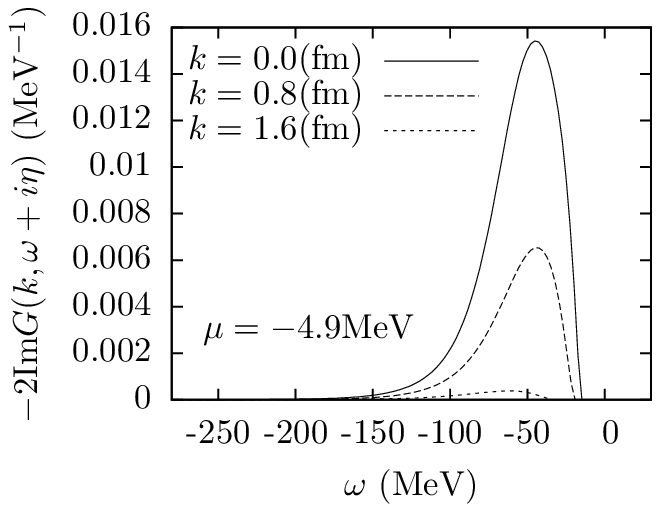}
\includegraphics[width=60mm]{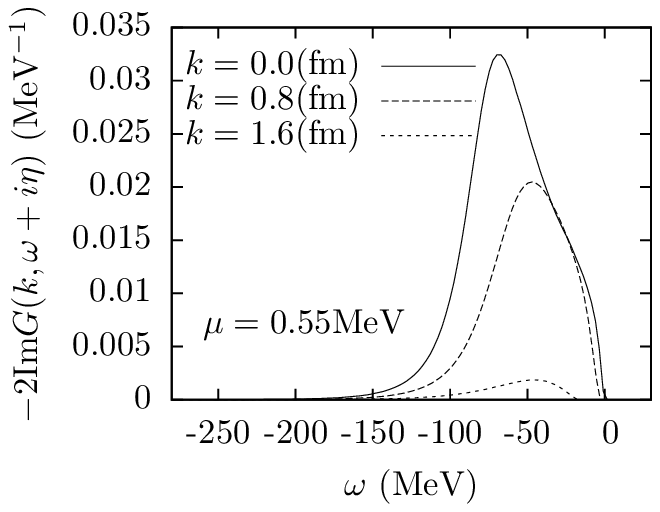}
\caption{
\label{fig-2ImG11}
$-2$Im$G(k,\omega+i\eta)$ in Eq.~(\ref{eq-Qmassoperator}) 
as function of $\omega$ for $\mu=-4.9$~MeV (top) and for $\mu=0.55$~MeV 
(bottom) at a zero temperature.\cite{Sog10}}
\end{center}
\end{figure}
 
The situation in the quartet case is also in so far much different, as the 
{$3h$} Green's function produces a considerable imaginary part of the mass 
operator.

Figure~\ref{fig-immassoperator} shows the imaginary part of the approximate 
quartet mass operator of Eq.~(\ref{eq-Qmassoperator}) for $\mu<0$ and 
$\mu>0$. These large values of the damping rate imply a strong violation 
of the quasiparticle picture. In Fig.~\ref{fig-2ImG11} we show the spectral 
function of the single particle GF. Contrary to the pairing case with its 
sharp quasiparticle pole, we here only find a very broad distribution, 
implying that the quasiparticle picture is completely destroyed. 
How to formulate a theory which goes continuously from the quartet case 
into the pairing case is, as mentioned, 
an open question. One solution could be to start 
right from the beginning with an in medium four body equation which contains 
a superfluid phase. When the quartet phase disappears, the superfluid phase 
may remain. Such investigations shall be done in the future.

\section{\label{sec-4}
Discussion, perspective, and conculision}

\begin{figure}[t]
\begin{center}
\includegraphics[width=120mm]{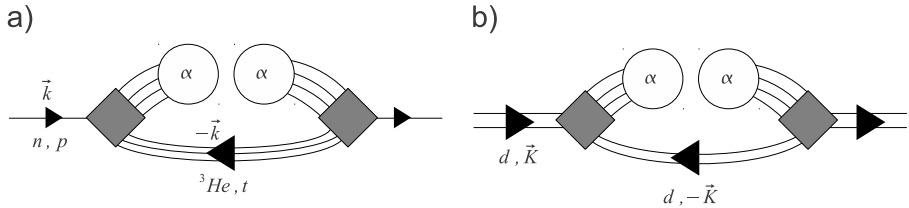}
\caption{(a) Schematic representation of single particle mass operator 
with $\alpha$ particle condensate pairing up a nucleon (n,p) with a 
trion ($^3$He, t); (b) mass operator for deuteron propagation pairing 
up with another deuteron of opposite momentum.}
\label{fig-10}
\end{center}
\end{figure}

The considerations in \S\ref{sec-3} are adequate for a situation where the 
uncorrelated Fermi gas directly goes over into the $\alpha$-particle 
condensed phase. However, in reality, in the low density phase of nuclear 
matter, the process may go via bound states of tritons or helions. 
One may imagine a mixture of deuterons, tritons, helions, and nucleons. 
The capture of 
a further nucleon of the triples can lead to $\alpha$ condensation. The mass 
operator of such a process is depicted in Fig.~\ref{fig-10}(a). 
Since the triples now have 
a definite c.o.m. momentum, stable quasi particles can form again. One has to 
describe such a process as with usual pairing but generalised to the case of 
two fermions with unequal masses.\cite{Orso} 
Another possible channel for $\alpha$ 
condensation is its formation out of two deuterons, see Fig.~\ref{fig-10}(b). 
This is 
similar to pairing 
of two bosons.\cite{schuck08} Which of 
the various 
processes will in the end win has to be evaluated in the future. Another open 
question is how the quartet order parameter behaves as a function of asymmetry 
and temperature. The dependence on asymmetry can partially be deduced from 
our study of the critical temperature in \S\ref{sec-3}. 
However, even the latter is 
incomplete, since only valid at the upper end of the density. In the limit as 
the density goes to zero we have to generalise the study of the critical 
temperature in such a way that the one of the ideal Bose gas is obtained. In 
the pairing case this has been achieved by Nozi\`eres and Schmitt-
Rink.\cite{Noz85} 
It is certainly a challenge to incorporate such an approach to quartet 
condensation. 
Coulomb repulsion between the $\alpha$-particles may
eventually favor an $\alpha$ crystal structure.\cite{takemoto}
Competition between Bose condensed phase and crystalization
may be investigated in future.
Quartetting is not confined to nuclear physics. As 
mentioned several times, there exists the possibility of the formation of a 
gas of bi-excitons in semiconductors.\cite{biexcitons} Also, one may speculate 
about the 
possibility that with cold atoms four different species of fermions will be 
trapped in the future. If they attract one another as the four nucleons 
in nuclear physics, 
certainly the quartet condensation can also be studied in such systems. First 
theoretical investigations have already appeared.\cite{Lecheminant05} 
The theory of 
quartetting in infinite systems is certainly not complete. More theoretical 
investigations are needed to get a full picture of the process.

\section*{Acknowledgements}
We greatfully acknowledge the very fruitful collaboration with our 
japanese collegues, 
Y. Funaki, H. Horiuchi, A. Tohsaki, T. Yamada on $\alpha$ particle 
condensation in finite nuclei.

%

\end{document}